# Evidence for a small hole pocket in the Fermi surface of underdoped YBa$_2$Cu$_3$O$_y$

N. Doiron-Leyraud[1], S. Badoux[2], S. René de Cotret[1], S. Lepault[2], D. LeBoeuf[2],
F. Laliberté[1], E. Hassinger[1], B. J. Ramshaw[3], D. A. Bonn[3,4], W. N. Hardy[3,4], R. Liang[3,4],
J.-H. Park[5], D. Vignolles[2], B. Vignolle[2], L. Taillefer[1,4] & C. Proust[2,4]

*1 Département de physique & RQMP, Université de Sherbrooke, Sherbrooke, Québec*
*J1K 2R1, Canada*

*2 Laboratoire National des Champs Magnétiques Intenses (CNRS, INSA, UJF, UPS),*
*Toulouse 31400, France*

*3 Department of Physics & Astronomy, University of British Columbia, Vancouver,*
*British Columbia V6T 1Z1, Canada*

*4 Canadian Institute for Advanced Research, Toronto, Ontario M5G 1Z8, Canada*

*5 National High Magnetic Field Laboratory, Tallahassee, Florida, USA*

Correspondence and requests for materials should be addressed to :
L.T. (email: louis.taillefer@usherbrooke.ca) or C.P. (email: cyril.proust@lncmi.cnrs.fr).

**In underdoped cuprate superconductors, the Fermi surface undergoes a reconstruction that produces a small electron pocket, but whether there is another, as yet undetected portion to the Fermi surface is unknown. Establishing the complete topology of the Fermi surface is key to identifying the mechanism responsible for its reconstruction. Here we report evidence for a second Fermi pocket in underdoped YBa$_2$Cu$_3$O$_y$, detected as a small quantum oscillation frequency in the thermoelectric response and in the $c$-axis resistance. The field-**

**angle dependence of the frequency shows that it is a distinct Fermi surface, and the normal-state thermopower requires it to be a hole pocket. A Fermi surface consisting of one electron pocket and two hole pockets with the measured areas and masses is consistent with a Fermi-surface reconstruction by the charge-density-wave order observed in $YBa_2Cu_3O_y$, provided other parts of the reconstructed Fermi surface are removed by a separate mechanism, possibly the pseudogap.**

## Introduction

The phase diagram of cuprate superconductors is shaped by ordered states and their identification is essential for understanding high temperature superconductivity. Evidence for a new state with broken symmetry in cuprates recently came from two major developments. The observation of quantum oscillations in underdoped $YBa_2Cu_3O_y$ (YBCO) (Ref. 1) and $HgBa_2CuO_{4+d}$ (Hg1201) (Ref. 2), combined with negative Hall[3,4,5] and Seebeck[5,6,7] coefficients, showed that the Fermi surface contains a small closed electron pocket and is therefore reconstructed at low temperature, implying that translational symmetry is broken. The detailed similarity of the Fermi-surface reconstruction in YBCO and $La_{1.6-x}Eu_{0.4}Sr_xCuO_4$ (Eu-LSCO) (Refs. 6,7,8) revealed that YBCO must host a density-wave order similar to the stripe order of Eu-LSCO (Ref. 9). More recently, charge-density-wave (CDW) modulations were observed directly, first by NMR in YBCO (Ref. 10) and then by x-ray diffraction in YBCO (Refs. 11,12,13) and Hg1201 (Ref. 14). In YBCO, a thermodynamic signature of the CDW order was detected in the sound velocity at low temperature and finite magnetic field[15]. These CDW modulations are reminiscent of the checkerboard pattern observed by STM on $Bi_2Sr_2CaCu_2O_{8+d}$ (Refs. 16,17), for instance.

Fermi-surface reconstruction and CDW modulations are therefore two universal signatures of underdoped cuprates, which begs the following question: Is the Fermi surface seen by quantum oscillations compatible with a reconstruction by the observed CDW modulations ? This issue requires a detailed knowledge of the Fermi surface, to be compared with Fermi surface calculations based on the measured parameters of the CDW order, in the same material at the same doping. In this article we report quantum oscillations and thermopower measurements that reveal an additional, hole-like Fermi pocket in underdoped YBCO. As we discuss below, a Fermi surface consisting of one electron and two hole pockets of the measured sizes and masses is consistent with a reconstruction by the observed CDW.

## Results

We have measured quantum oscillations in the thermoelectric response and $c$-axis resistance of underdoped YBCO. Our samples were chosen to have a doping $p = 0.11$-$0.12$, at which the amplitude of quantum oscillations is maximal[18]. In the doping phase diagram, this is also where the CDW modulations are strongest[19,20] (Fig. 1a) and where the critical field $H_{c2}$ needed to suppress superconductivity is at a local minimum[21] (Fig. 1b). In the $T = 0$ limit, the Seebeck ($S$) and Nernst ($v = N / B$) coefficients are inversely proportional to the Fermi energy[22,23], and are therefore expected to be enhanced for small Fermi surfaces. In Fig. 2a, we show isotherms of $S$ and $N$ at $T = 2$ K measured up to $B = 45$ T in a YBCO sample with $p = 0.11$. Above $H_{c2} = 24$ T (Ref. 21), both $S$ and $N$ are negative ; the fact that $S < 0$ is consistent with an electron pocket dominating the transport at low temperature[6,7]. The normal-state signal displays exceptionally large quantum oscillations, with a main frequency $F_a = 540$ T and a beat pattern indicative of other, nearby, frequencies. In Fig. 2, we also show the $c$-axis resistance of two YBCO samples at $p = 0.11$ and $0.12$, measured in pulsed fields up to 68 T. The overall behaviour of the $c$-axis magnetoresistance at $p = 0.11$ is consistent with previous reports[24,25]. Quantum oscillations are clearly visible and the three distinct frequencies $F_{a1} = 540$ T, $F_{a2} = 450$ T and $F_{a3} = 630$ T in the Fourier spectrum at $p = 0.11$ (Fig. 1d) agree with reported values[26].

With increasing temperature, the amplitude of these "fast" oscillations decreases rapidly and above $T \sim 10$ K we are left with a slowly undulating normal-state signal, clearly seen in the raw Seebeck data (Figs. 2b and 3a). In Fig. 3b, the oscillatory part of that signal, obtained by subtracting a smooth background, is plotted as a function of inverse magnetic field. Although the "slow oscillations" at 18 K are 20 times weaker than the fast oscillations at 2 K, they are clearly resolved and periodic in $1 / B$. After their discovery in the Seebeck signal, the slow oscillations were also detected in the $c$-axis resistance, as shown in Fig. 3c. In both the Seebeck and $c$-axis resistance data, the frequency of these slow oscillations is $F_b = 95 \pm 10$ T ($p = 0.11$). Similar oscillations were also detected in the $c$-axis resistance of a sample at $p = 0.12$ (Fig. 3d), with $F_b = 120 \pm 15$ T. In Fig. 4a, we show the derivative $dR_c / dB$, which unambiguously reveals $F_b$, without the need for a background subtraction. (Note that in the $c$-axis resistance data, the amplitude of $F_b$ is about 0.1% of the total signal and is more sensitive to the background subtraction.) This slow frequency persists up to 30 K and its amplitude follows the usual Lifshitz-Kosevich formula (Fig. 4b), with a small effective mass $m^* = 0.45 \pm 0.1\, m_0$, where $m_0$ is the free electron mass.

Using the $c$-axis resistance, we have measured the dependence of $F_b$ on the angle $\theta$ at which the field is tilted away from the $c$ axis. In Fig. 4c, the oscillatory part of the $c$-axis resistance for $p = 0.11$ at $T = 15$ K is plotted versus $1 / B\cos(\theta)$, and the angular dependence of $F_b$ is displayed in Fig. 4d. $F_b(\theta)$ varies approximately as $1 / \cos(\theta)$, indicating that the Fermi surface associated with $F_b$ is a warped cylinder along the $c$ axis, as expected for a quasi-2D system.

## Discussion

The slow frequency $F_b$ ~ 100 T reported here bears the key signatures of quantum oscillations and in the following discussion we argue that it comes from a small hole-like Fermi surface, distinct from the larger electron-like Fermi pocket responsible for the main frequency $F_{a1} = 540$ T.

We note that the frequency $F_b$ is nearly equal to the difference between the main frequency of the electron pocket $F_{a1}$ and its satellites $F_{a2}$ and $F_{a3}$. While the identification of the multiple $F_a$ frequencies is not definitive, it is likely that two of them are associated with the two separate Fermi surfaces that come from the two $CuO_2$ planes (bilayer) in the unit cell of YBCO. The third frequency could then either come from magnetic breakdown between these two Fermi surfaces[27] or from a warping due to $c$-axis dispersion[26,28]. In layered quasi-2D materials, slow quantum oscillations can appear in the $c$-axis transport as a result of interlayer coupling[29,30]. Two observations allow us to rule out this scenario in the present context. First, $F_b$ is observed in the in-plane Seebeck coefficient, which does not depend on the $c$-axis conductivity. Secondly, at a special field angle $\theta$, called the Yamaji angle, where the $c$-axis velocity vanishes on average along a cyclotron orbit, one should see a vanishing $F_b$. This is not seen in our field-angle dependence of $F_b$, which, if anything, only deviates upward from a cylindrical $1 / B\cos(\theta)$ dependence (Fig. 4d).

Quantum interference from magnetic breakdown between two bilayer-split orbits could in principle produce a difference frequency close to $F_b$. In this scenario, however, the amplitudes of the two nearby frequencies $F_{a2}$ and $F_{a3}$ should be identical, irrespective of the field range, in disagreement with torque[26] and $c$-axis resistance measurements (see Fig. 1d). Furthermore, in a magnetic breakdown scenario, we would expect $F_b$ to scale with $F_a$, since both frequencies originate from the same cyclotron orbits. This is not what we observe in our thermoelectric data: as seen in Fig. 2a, the amplitude of $F_a$ is larger in the Nernst effect than in the Seebeck effect, yet $F_b$ is only detected in the latter. This is strong evidence that $F_a$ and $F_b$ do not involve cyclotron orbits on the same Fermi surface. We therefore conclude that $F_b$ must come from a distinct Fermi pocket, in contrast with the interpretation of Ref. 31 in terms of quantum interference.

For a number of reasons, we infer that this second pocket in the reconstructed Fermi surface of YBCO is hole-like. The first reason is the strong dependence of resistivity $\rho$ and Hall coefficient $R_H$ on magnetic field $B$, as observed in YBCO and in $YBa_2Cu_4O_8$ (Ref. 3), a closely related material with similar quantum oscillations[32,33]. For instance, $R_H(B)$ goes from positive at low field to negative at high field[3] and $\rho(B)$ exhibits a significant magnetoresistance[25]. These are natural consequence of having both electron and hole carriers. In $YBa_2Cu_4O_8$, the Hall and resistivity data were successfully fit in detail to a two-band model of electrons and holes[34].

A second indication that both electron and hole carriers are present in underdoped YBCO is the fact that quantum oscillations are observed in the Hall coefficient[35,36]. In an isotropic single-band model, the Hall coefficient is simply given by $R_H = 1 / ne$, where $n$ is the carrier density and $e$ the electron charge. Quantum oscillations in $R_H$ appear via the scattering rate, which enters $R_H$ either when two or more bands of different mobility are present, or when the scattering rate on a single band is strongly anisotropic. At low temperatures, however, where impurity scattering dominates, the latter scenario is improbable.

The most compelling evidence for the presence of hole-like carriers in underdoped YBCO comes from the magnitude of the Seebeck coefficient. In the $T = 0$ limit and for a single band, it is given by[22]:

$$\frac{S}{T} = \pm \frac{\pi^2}{3} \frac{k_B}{e} \frac{1}{T_F} \left( \frac{3}{2} + \zeta \right) \qquad \qquad ( \ 1 \ )$$

where $k_B$ is Boltzmann's constant, $T_F$ is the Fermi temperature, and $\zeta = 0$ or -1/2 depending on whether the relaxation time or the mean free path is assumed to be energy-independent, respectively. The sign of $S / T$ depends on whether the carriers are holes (+) or electrons (-). This expression has been found to work very well in a variety of correlated electron metals[22]. We stress that $S / T$ (in the $T = 0$ limit) is governed solely by $T_F$, which allows a direct quantitative comparison with quantum oscillation data, with no assumption on pocket multiplicity. This contrasts with the specific heat, which depends on the number of Fermi pockets (see below).

In Fig. 5a, we reproduce normal-state Seebeck data in YBCO at four dopings, plotted as $S / T$ vs $T$ (from Ref. 7). $S / T$ goes from positive at high $T$ to negative at low $T$, in agreement with a similar sign change in $R_H(T)$ (Refs. 3,4), both evidence that the dominant carriers at low $T$ are electron-like. Extrapolating $S / T$ to $T = 0$ as shown by the dashed lines in Fig. 5a, we obtain the residual values and plot them as a function of doping in Fig. 5b ($S_{measured}$, red squares). We see that the size of the residual term is largest (*i.e.*, is most strongly negative) at $p = 0.11$, and that it decreases on both sides.

This doping-dependent $S / T$ is to be compared with the Fermi temperature directly measured by quantum oscillations, via:

$$T_F = \frac{e\hbar}{k_B} \frac{F}{m^*} \qquad\qquad ( \; 2 \; )$$

where $F$ is the frequency, $m^*$ the effective mass, and assuming a parabolic dispersion. For YBCO at $p = 0.11$, the electron pocket gives $F_{a1} = 540 \pm 20$ T and $m^* = 1.76\, m_0$, so that $T_F = 410 \pm 20$ K and hence $S_e / T = -1.0$ $\mu$V / K$^2$ ($-0.7$ $\mu$V / K$^2$), for $\zeta = 0$ ($-1/2$). In Fig. 5a, the measured $S / T$ extrapolated to $T \rightarrow 0$ gives a value of $-0.9$ $\mu$V / K$^2$. The electron pocket alone therefore accounts by itself for essentially the entire measured Seebeck signal at $p = 0.11$. From quantum oscillation measurements at different dopings[24,37,38], we know the values of $F_{a1}$ and $m^*$ from $p = 0.09$ to $p = 0.13$, and can therefore determine the evolution of $S_e / T$ in that doping interval. The result is plotted as blue circles in Fig. 5b, where we see that the calculated $| S_e / T |$ increases by a factor 2.5 between $p = 0.11$ and $p = 0.09$. This is because the mass $m*$ increases strongly as $p \to 0.08$ (Ref. 38), while $F_{a1}$ decreases only slightly[18]. This strong *increase* in the calculated $| S_e / T |$ is in stark contrast with the measured value of $| S / T |$, which *decreases* by a factor 3 between $p = 0.11$ and $p = 0.09$ (Fig. 5b). To account for the observed doping dependence of the thermopower in YBCO, we are lead to conclude that there must be a hole-like contribution to $S / T$. We emphasize that the Hall coefficient $R_H$ (Ref. 4) measured well above $H_{c2}$ (Ref. 21) displays the same dome-like dependence on doping as the Seebeck coefficient (Ref. 7), which further confirms the presence of a hole-like Fermi pocket.

In a two-band model, the total Seebeck coefficient is given by

$$S = \frac{S_e \sigma_e + S_h \sigma_h}{\sigma_e + \sigma_h} \qquad\qquad ( \, 3 \, )$$

where the hole ($h$) and electron ($e$) contributions are weighted by their respective conductivities $\sigma_h$ and $\sigma_e$. As shown in Fig. 5c, we can account for the measured $S / T$ at $T \to 0$ by adding a hole-like contribution, $S_h$, which we estimate from $F_b = 95 \pm 10$ T and $m* = 0.45 \, m_0$, giving $T_F = 280 \pm 80$ K. Assuming for simplicity that $S_h$ is doping-independent leaves the ratio of conductivities, $\sigma_e / \sigma_h$, as the only adjustable parameter in the above two-band expression (Eq. 3). In Fig. 5d, we plot the resulting $\sigma_e / \sigma_h$ as a function of doping, and we see that it peaks at $p = 0.11$ and drops on either side. This is consistent with the fact that the amplitude of the fast quantum oscillations is largest at $p = 0.11$, and much smaller away from that doping[18], direct evidence that the mobility of the electron pocket is maximal at $p = 0.11$. This is clearly seen in the resistance data in Figs. 2c and 2d, where the amplitude of the quantum oscillations of the electron pocket is strongly reduced when going from $p = 0.11$ to 0.12. In contrast, Figs. 3c and 3d show that the amplitude of the oscillations from the hole pocket remains nearly constant: at $T = 4.2$ K and $H = 68$ T, their relative amplitude is $\Delta R_c / R_c = 0.036$ % at $p = 0.11$ and 0.03 % at $p = 0.12$. The change in conductivity ratio therefore comes mostly from a change in $\sigma_e$.

To summarize, in addition to the two-band description of transport data in YBa$_2$Cu$_4$O$_8$ (Ref. 34), the doping dependence of the Seebeck[7] and Hall[4] coefficients in YBCO is firm evidence that the reconstructed Fermi surface of underdoped YBCO (for $0.08 < p < 0.18$) contains not only the well-established electron pocket[4], but also another hole-like surface (of lower mobility). We combine this evidence with our discovery of an additional small Fermi surface to conclude that this new pocket is hole-like.

From the measured effective mass $m^*$, the residual linear term $\gamma$ in the electronic specific heat $C_e(T)$ at $T \rightarrow 0$ can be estimated through the relation (Ref. 38)

$$\gamma = (1.46 \text{ mJ} / \text{K}^2 \text{ mol}) \, \Sigma_i \, (n_i \, m_i^* / m_0) \qquad (4)$$

where $n_i$ is the multiplicity of the $i^{th}$ type of pocket in the first Brillouin zone (This expression assumes an isotropic Fermi liquid in two dimensions with a parabolic dispersion). For a Fermi surface containing one electron pocket and two hole pockets per CuO$_2$ plane, we obtain a total mass of $(1.7 \pm 0.2) + 2 \, (0.45 \pm 0.1) = 2.6 \pm 0.4 \, m_0$, giving $\gamma = 7.6 \pm 0.8$ mJ / K$^2$ mol (for two CuO$_2$ planes per unit cell). High-field measurements of $C_e$ at $T \rightarrow 0$ in YBCO at $p \sim 0.1$ yield $\gamma = 5 \pm 1$ mJ / K$^2$ mol (Ref. 39) at $H > H_{c2} = 30$ T (Ref. 21). We therefore find that the Fermi surface of YBCO can contain at most two of the small hole pockets reported here, in addition to only one electron pocket. No further sheet can realistically be present in the Fermi surface.

There is compelling evidence that the Fermi surface of YBCO is reconstructed by the CDW order detected by x-ray diffraction. In particular, Fermi-surface reconstruction[4,7] and CDW modulations[19,20] are detected in precisely the same region of the temperature-doping phase diagram. Because the CDW modulations are along both the *a* and *b* axes, the reconstruction naturally produces a small closed electron pocket along the Brillouin zone diagonal, at the so-called nodal position[40,41]. Given the wavevectors measured by x-ray diffraction, there will also be small closed hole-like ellipses located between the diamond-shaped nodal electron pockets. An example of the Fermi surface calculated[42] using the measured CDW wavevectors is sketched in Fig. 1c. It contains two distinct closed pockets: a nodal electron pocket of area such that $F_e \sim 500$ T and a hole-like ellipse such that $F_h \sim 100$ T (Ref. 42). Note that a reconstruction by a commensurate wavevector $q = 1/3\ \pi/a$, very close to the measured value, yields one electron and two hole pockets per Brillouin zone, as assumed in our calculation of $\gamma$ above.

If, as indeed observed in YBCO at $p = 0.11$ (Ref. 20), the CDW modulations are anisotropic in the *a-b* plane, the ellipse pointing along the *a* axis will be different from that pointing along the *b* axis[42]. If one of the ellipses is close enough to the electron pocket, *i.e.* if the gap between the two is small enough, magnetic breakdown will occur between the hole and the electron pockets, and this could explain the complex spectra of multiple quantum oscillations seen in underdoped YBCO (Fig. 1d).

In most models of Fermi-surface reconstruction by CDW order, the size of the Fermi pockets can be made to agree with experiments using a reasonable set of parameters. For example, a similar Fermi surface (with one electron pocket and two hole pockets) is also obtained if one considers a « criss-crossed » stripe pattern instead of a checkerboard[43]. At this level, there is consistency between our quantum oscillation measurements and models of Fermi-surface reconstruction by the CDW order. However, in addition to the electron and hole pockets, the folding of the large Fermi surface produces other segments of Fermi surface whose total contribution to $\gamma$ greatly exceeds that allowed by the specific heat data. Consequently, there must exist a mechanism that removes parts of the Fermi surface beyond the reconstruction by the CDW order. A possible mechanism is the pseudogap. The loss of the anti-nodal states caused by the pseudogap would certainly remove parts of the reconstructed Fermi surface.

Further theoretical investigations are needed to understand how pseudogap and CDW order are intertwined in underdoped cuprates. An important point in this respect is the fact that, unlike in $Bi_2Sr_{2-x}La_xCuO_{6+\delta}$ (Ref. 44), the CDW wavevector measured by x-ray diffraction in YBCO and in Hg1201 (Ref. 14) does not connect the hot spots where the large Fermi surface intersects the antiferromagnetic Brillouin zone (Fig. 1c), nor does it nest the flat anti-nodal parts of that large Fermi surface (Fig. 1c).

## Methods

**Samples.** Single crystals of $YBa_2Cu_3O_y$ (YBCO) with $y = 6.54$, 6.62, and 6.67 were obtained by flux growth at UBC (Ref. 46). The superconducting transition temperature $T_c$ was determined as the temperature below which the zero-field resistance $R = 0$. The hole doping $p$ is obtained from $T_c$ (Ref. 47), giving $p = 0.11$ for $y = 6.54$ and 6.62, and $p = 0.12$ for $y = 6.67$. The samples with $y = 6.54$ and 6.62 have a high degree of ortho-II oxygen order, and the sample with $y = 6.67$ has ortho-VIII order. The samples are detwinned rectangular platelets, with the $a$ axis parallel to the length (longest dimension) and the $b$ axis parallel to the width. The electrical contacts are diffused evaporated gold pads with a contact resistance less than 1 $\Omega$.

**Thermoelectric measurements.** The thermoelectric response of YBCO with $y = 6.54$ ($p = 0.11$) was measured at the National High Magnetic Field Laboratory (NHMFL) in Tallahassee, Florida, up to 45 T, in the temperature range from 2 K to 40 K. The Seebeck and Nernst coefficients are given by $S \equiv -\nabla V_x / \nabla T_x$ and $\nu \equiv N / B \equiv (\nabla V_y / \nabla T_x) / B$, respectively, where $\nabla V_x$ ($\nabla V_y$) is the longitudinal (transverse) voltage gradient caused by a temperature gradient $\nabla T_x$, in a magnetic field $\boldsymbol{B} \parallel \boldsymbol{z}$. A constant heat current was sent along the $a$ axis of the single crystal, generating a temperature difference $\Delta T_x$ across the sample. $\Delta T_x$ was measured with two uncalibrated Cernox chip thermometers (Lakeshore), referenced to a third, calibrated Cernox. The longitudinal and transverse electric fields were measured using nanovolt preamplifiers and nanovoltmeters. All measurements were performed with the temperature of the experiment stabilized within $\pm$ 10 mK and the magnetic field $B$ swept at a constant rate of 0.4 – 0.9 T / min between positive and negative maximal values, with the heat on. The field was applied normal to the CuO$_2$ planes ($B \parallel z \parallel c$).

Since the Seebeck coefficient $S$ is symmetric with respect to the magnetic field, it is obtained by taking the mean value between positive and negative fields:

$$S = E_x / ( \partial T / \partial x ) = [ \Delta V_x(B) + \Delta V_x(-B) ] / ( 2 \Delta T_x ) \quad,$$

where $\Delta V_x$ is the difference in the voltage along $x$ measured with and without thermal gradient. This procedure removes any transverse contribution that could appear due to slightly misaligned contacts. The longitudinal voltages and the thermal gradient being measured on the same pair of contacts, no geometric factor is involved.

The Nernst coefficient $N$ is anti-symmetric with respect to the magnetic field, therefore it is obtained by the difference:

$$N = E_y / ( \partial T / \partial x ) = ( L / w ) [ V_y(B) - V_y(-B) ] / ( 2 \Delta T_x ) \quad,$$

where $L$ and $w$ are the length and width of the sample, respectively along $x$ and $y$, and $V_y$ is the voltage along $y$ measured with the heat current on. This anti-symmetrisation procedure removes any longitudinal thermoelectric contribution and a constant background from the measurement circuit. The uncertainty on $N$ comes from the uncertainty in determining $L$ and $w$, giving typically an error bar of $\pm$ 10 %.

**Resistance measurements.** The *c*-axis resistance was measured at the Laboratoire National des Champs Magnétiques Intenses (LNCMI) in Toulouse, France, in pulsed magnetic fields up to 68 T. Measurements were performed in a conventional four-point configuration, with a current excitation of 5 mA at a frequency of ~ 60 kHz. Electrical contacts to the sample were made with large current pads and small voltage pads mounted across the top and bottom so as to short out any in-plane current. A high-speed acquisition system was used to digitize the reference signal (current) and the voltage drop across the sample at a frequency of 500 kHz. The data were analyzed with software that performs the phase comparison. $\theta$ is the angle between the magnetic field and the *c* axis, and measurements were done at $\theta = 0°$ up to 68 T and at various angles $\theta$ up to 58 T. The uncertainty on the absolute value of the angle is about 1°.

## Acknowledgements

We thank A. Allais, K. Behnia, A. Carrington, S. Chakravarty, B. Fauqué, A. Georges, N. E. Hussey, M.-H. Julien, S.A. Kivelson, M.R. Norman, S. Raghu, S. Sachdev, A.-M. Tremblay, and C. Varma for fruitful discussions. We thank J. Béard and P. Frings for their assistance with the experiments at the LNCMI. The work in Toulouse was supported by the French ANR SUPERFIELD, the EMFL, and the LABEX NEXT. A portion of this work was performed at the National High Magnetic Field Laboratory, which is supported by the National Science Foundation Cooperative Agreement No. DMR-1157490, the State of Florida, and the U.S. Department of Energy. R.L., D.A.B. and W.N.H. acknowledge support from NSERC. L.T. acknowledges support from the Canadian Institute for Advanced Research and funding from NSERC, FRQNT, the Canada Foundation for Innovation, and a Canada Research Chair.

## Author contributions

N.D.-L., S.R.d.C. and J.-H.P. performed the Seebeck and Nernst measurements at the NHMFL in Tallahassee. N.D.-L., F.L. and E.H. analyzed the Seebeck data. S.B., S.L., D.L., D.V., B.V. and C.P. performed and analyzed the resistance measurements at the LNCMI in Toulouse. B.J.R., R.L., D.A.B. and W.N.H. prepared the YBCO single crystals at UBC (crystal growth, annealing, de-twinning, contacts). L.T. and N.D.-L. supervised the thermo-electric measurements. C.P. supervised the pulsed-field measurements. N.D.-L., L.T. and C.P. wrote the manuscript, with input from all authors.

### Figure 1 | Phase diagrams and Fermi surface of YBCO

**a)** Temperature-doping phase diagram of YBCO, showing the superconducting transition temperature $T_c$ (grey dome, Ref. 47) and the CDW onset temperature $T_{CDW}$ (green diamonds, Ref. 19; blue diamonds, Ref. 20). **b)** Upper critical field $H_{c2}$ of YBCO as a function of doping (red dots, Ref. 21). In both a) and b), the arrows indicate the two dopings of the samples used in our study. All error bars in panels a) and b) are reproduced from the original references. **c)** Sketch of the reconstructed Fermi surface adapted from Ref. 42 using the CDW wavevectors (arrows) measured in YBCO (Refs. 19, 20), showing a diamond-shaped nodal electron pocket (red) and two hole-like ellipses (blue and green). The dashed line is the antiferromagnetic Brillouin zone; the dotted line is the original large Fermi surface; the black dots mark the so-called "hot spots", where those two lines intersect. **d)** Fourier transform of our $c$-axis resistance data (at $p$ = 0.11), showing the new "low" frequency $F_b$ = 95 ± 10 T reported here, and the three main "high" frequencies $F_{a1}$ (blue), $F_{a2}$ (red), and $F_{a3}$ (green).

### Figure 2 | Quantum oscillations in YBCO

**a)** Seebeck ($S$; red) and Nernst ($N$; blue) signals in YBCO $p$ = 0.11 as a function of magnetic field $B$ at $T$ = 2 K. **b)** Seebeck coefficient $S$, plotted as $S / T$ vs $B$, at temperatures as indicated. **c, d)** $c$-axis electrical resistance $R_c$ of YBCO samples with $p$ = 0.11 and $p$ = 0.12, as a function of $B$ up to 68 T, at different temperatures as indicated. For all data, the field $B$ is along the $c$ axis.

**Figure 3 | Slow quantum oscillations**

**a)** Seebeck coefficient in YBCO $p = 0.11$ (blue), plotted as $S / T$, as a function of field $B$ at $T = 18$ K, showing slow oscillations about a linear background (red). **b)** Oscillatory part of the Seebeck coefficient $\Delta S / T$ (obtained by subtracting a $2^{nd}$ order polynomial from the raw data) as a function of $1 / B$, showing the usual fast quantum oscillations at $T = 2$ K (red), and the new slow oscillations with $F_b = 95 \pm 10$ T at $T = 18$ K (blue, multiplied by 20). For clarity, the slow frequency $F_b$ was removed from the data at $T = 2$ K. **c, d)** Oscillatory part of the $c$-axis electrical resistance $\Delta R_c$ in YBCO (obtained by subtracting a $3^{rd}$ order polynomial from the raw data) at $p = 0.11$ and $p = 0.12$ as a function of $1 / B$, at temperatures as indicated. The oscillations are periodic in $1 / B$, with a frequency $F_b = 95 \pm 10$ T and $120 \pm 15$ T at $p = 0.11$ and 0.12, respectively.

**Figure 4 | Properties of the slow frequency at $p = 0.11$**

**a)** Derivative of the resistance of sample $p = 0.11$ with respect to field $B$, plotted versus $1 / B$ at temperatures as indicated. This confirms the presence of the slow frequency $F_b$, irrespective of background subtraction. **b)** Amplitude of $F_b$ oscillation as a function of temperature (dots). The line is a Lifshitz-Kosevich fit to the data, giving an effective mass $m^* = 0.45 \pm 0.1 \; m_0$. **c)** Oscillatory part of the $c$-axis resistance at different angles $\theta$ between the field and the $c$ axis, as indicated, as a function of $1 / B\cos(\theta)$ at $T = 15$ K. A $2^{nd}$ order polynomial background was subtracted from the raw data to extract $\Delta R$. **d)** Slow frequency $F_b$ as a function of $\theta$ (dots). The red line is the function $1 / \cos(\theta)$. The error bars are a convolution of standard deviations in the value of $F$, for different fitting ranges and different orders of the polynomial background.

**Figure 5 | Evidence of a hole-like contribution to the Seebeck coefficient**

**a)** Normal-state Seebeck coefficient $S / T$ as a function of temperature for YBCO at four dopings, as indicated (adapted from Ref. 7). The squares and dots are data in zero field and in 28 T, respectively. The dotted lines are extrapolations of $S / T$ to $T \rightarrow 0$, whose values are plotted in b) and c) (red squares). The lines are a guide to the eye. **b)** Extrapolated value of $S / T$ at $T = 0$ as a function of doping ($S_{\text{measured}}$, red squares; from data and extrapolations in a)). The error bars represent the uncertainty in extrapolating to $T = 0$. The blue dots ($S_e$) indicate $S / T$ ($T \rightarrow 0$) for the electron pocket alone, calculated from the main quantum oscillation frequency $F_{a1}$ and mass $m^*$ (from Refs. 24, 37, 38) (using Eqs. 1 and 2). The blue line is a guide to the eye. **c)** The red squares ($S_{\text{measured}}$) are identical to those in b). Using a two-band model (Eq. 3), we include the contribution of the hole pocket ($S_h$) using the measured Fermi temperature associated with the slow frequency $F_b$ ($T_F = 280 \pm 80$ K; see text). With the conductivity ratio $\sigma_e / \sigma_h$ as the only fit parameter, this model (black line) reproduces the measured data (red squares). **d)** Ratio of electron to hole conductivities, $\sigma_e / \sigma_h$ (black circles), from a fit (black line, panel c)) to the measured values of $S / T$ at $T = 0$ (red squares, panels b) and c)). The black line is a guide to the eye. The error bars are derived from the errors on $S_{\text{measured}}$.

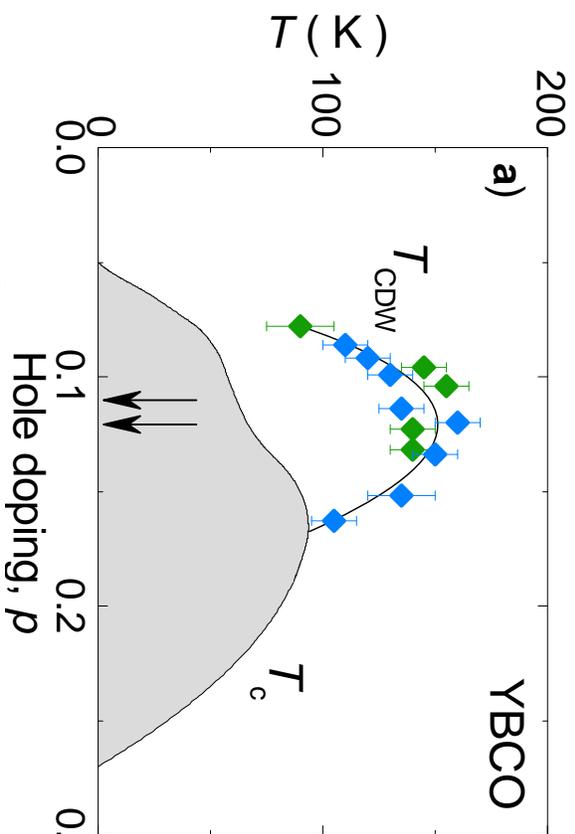

**a)** YBCO

$T$ (K)

$T_{CDW}$

$T_c$

Hole doping, $p$

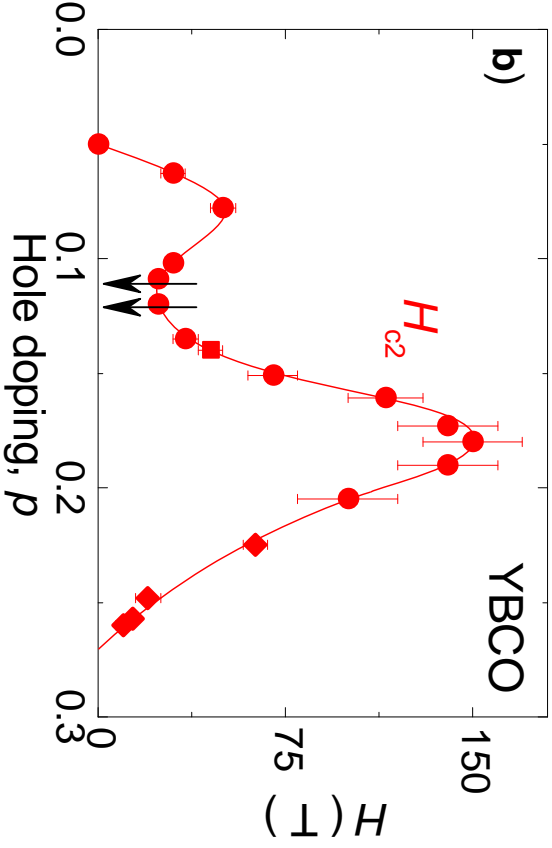

**b)** YBCO

$H_{c2}$

$H$ (T)

Hole doping, $p$

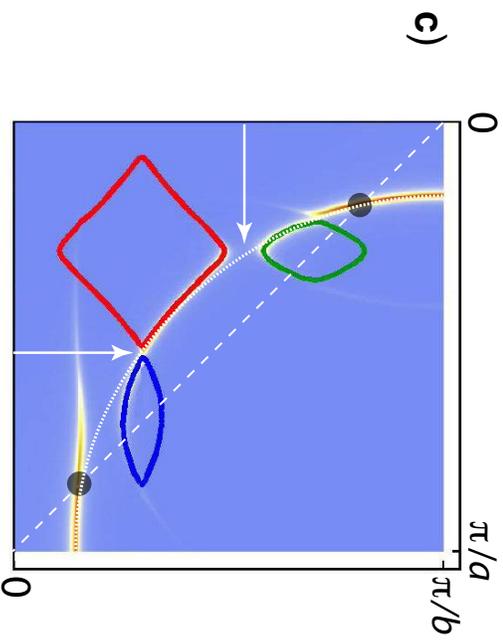

**c)**

$\pi/a$

$\pi/b$

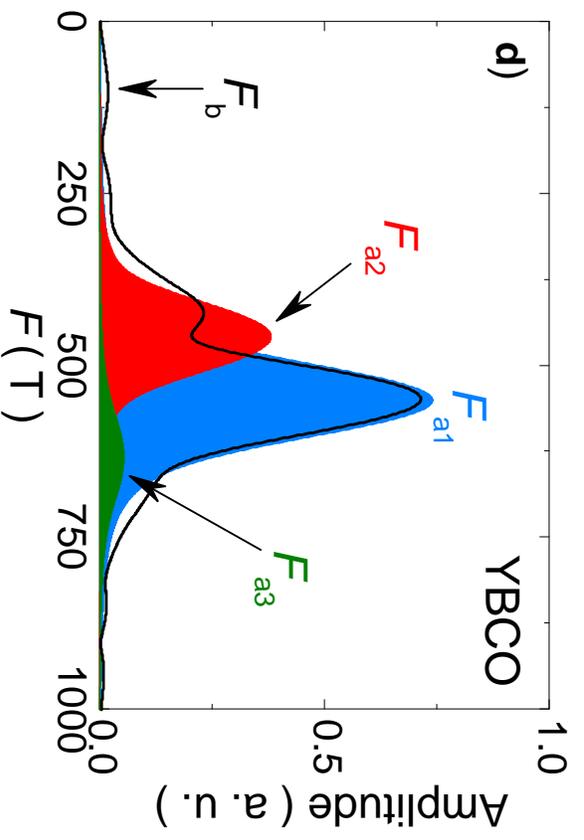

**d)** YBCO

$F_b$

$F_{a2}$

$F_{a1}$

$F_{a3}$

Amplitude (a.u.)

$F$ (T)

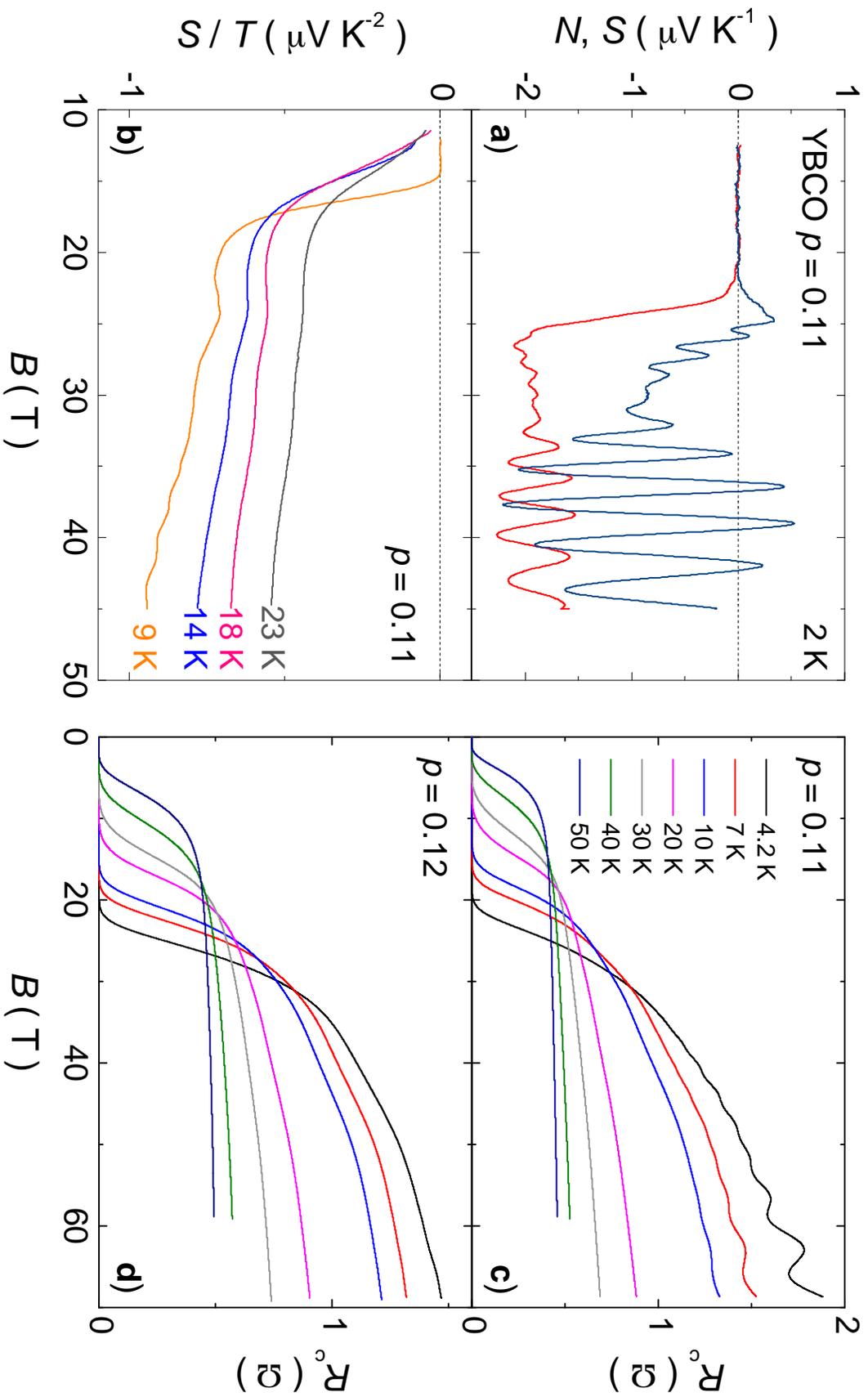

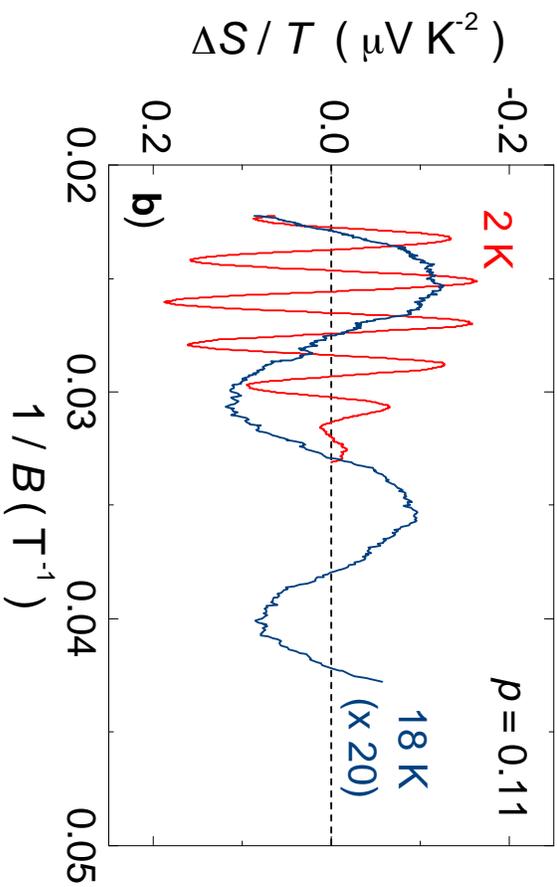

**a)** YBCO $p = 0.11$

$S / T$ ($\mu$V K$^{-2}$)

$T = 18$ K

$B$ (T)

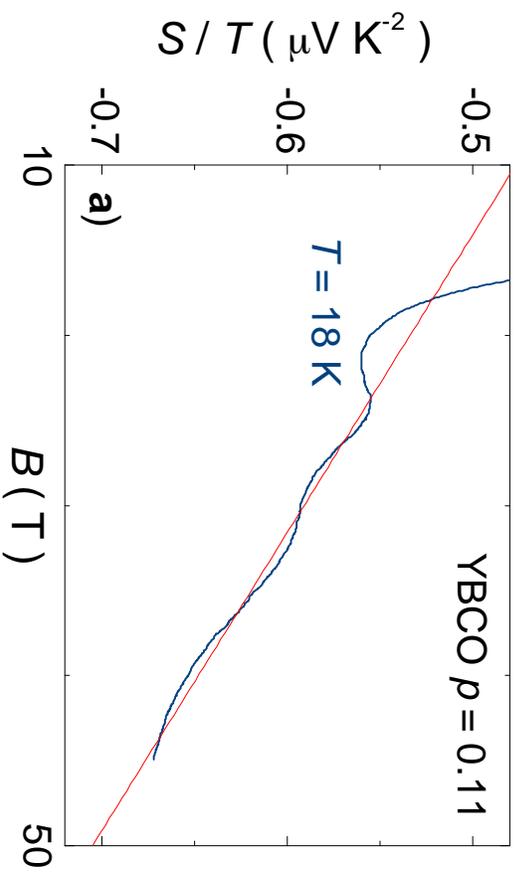

**c)** $p = 0.11$

$\Delta R_c$ (m$\Omega$)

$1 / B$ (T$^{-1}$)

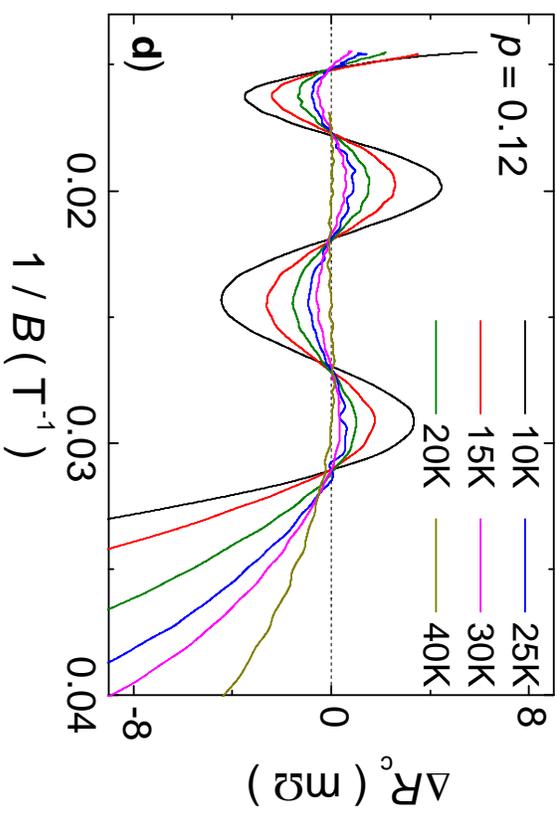

**b)** $p = 0.11$

$\Delta S / T$ ($\mu$V K$^{-2}$)

2 K

18 K (x 20)

$1 / B$ (T$^{-1}$)

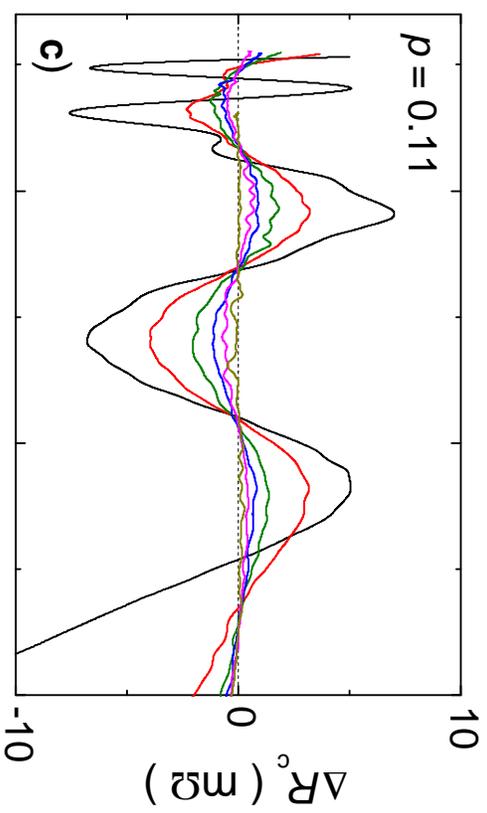

**d)** $p = 0.12$

$\Delta R_c$ (m$\Omega$)

$1 / B$ (T$^{-1}$)

- 10K
- 15K
- 20K
- 25K
- 30K
- 40K

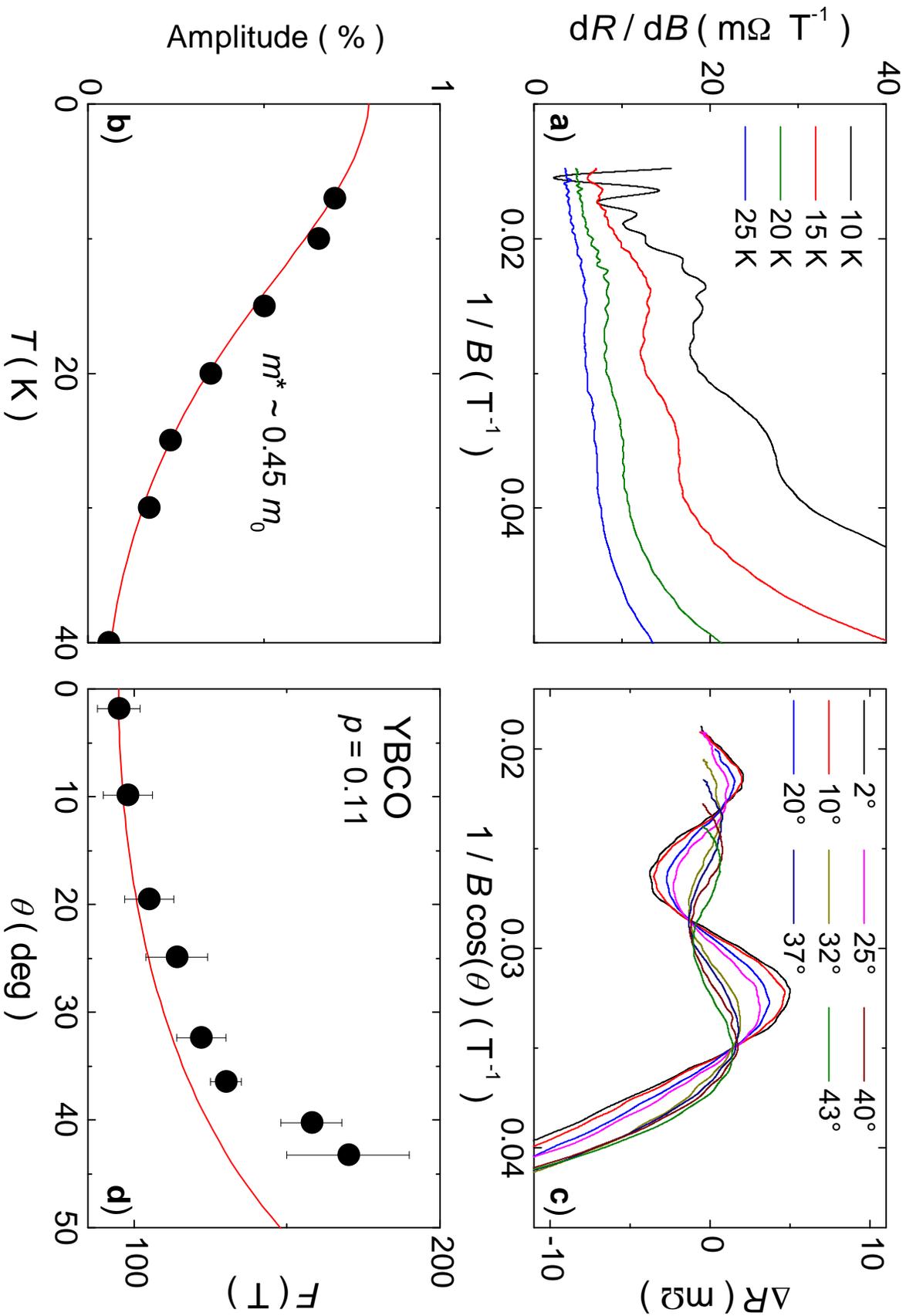

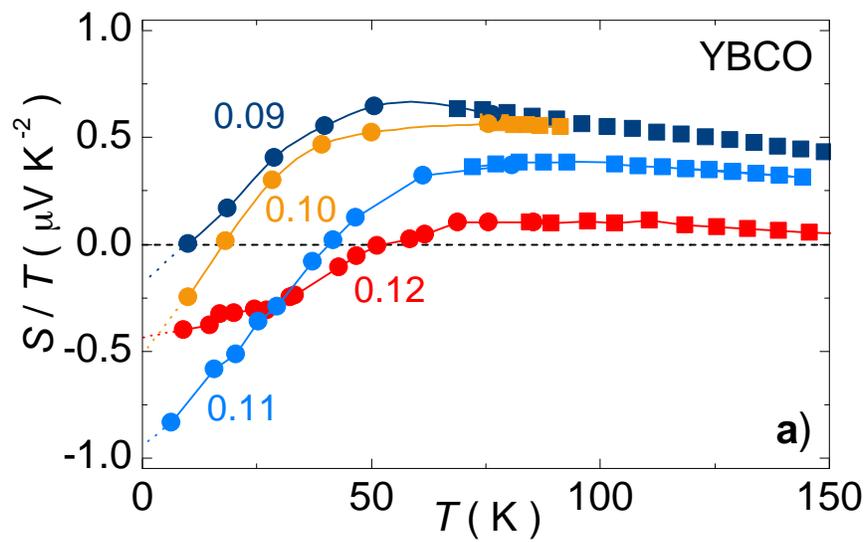

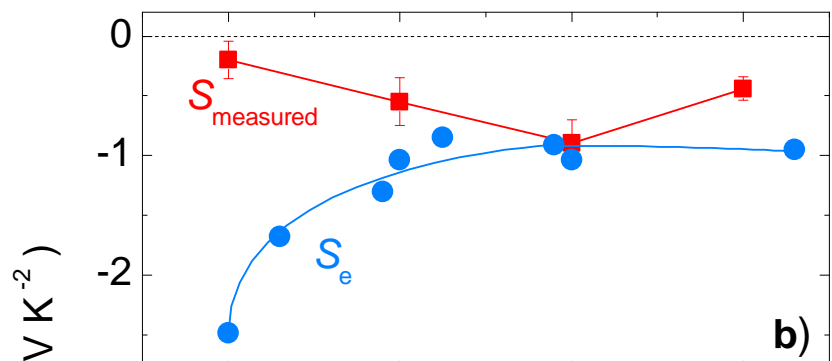

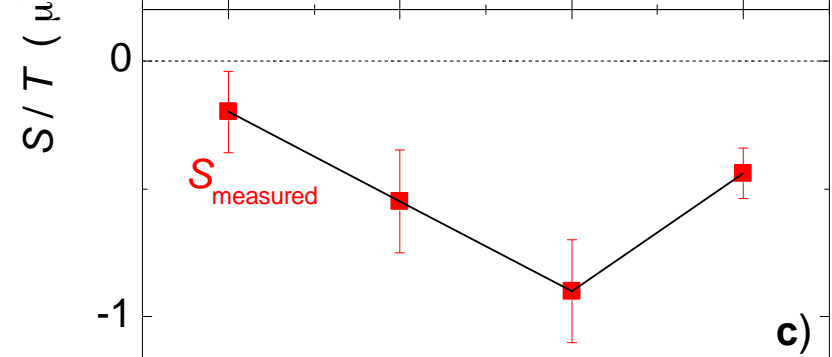

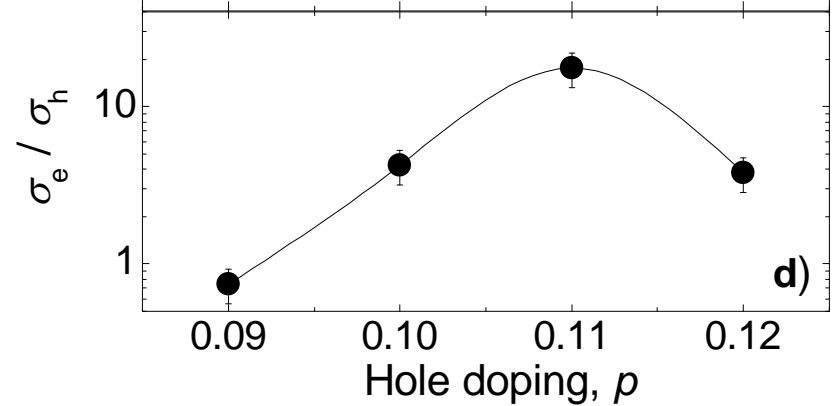